\documentclass[a4paper,11pt]{article}
\pdfoutput=1 % if your are submitting a pdflatex (i.e. if you have
\usepackage{jinstpub} % for details on the use of the package, please
\usepackage{mathrsfs}

\title{Wakefields of the FCC-ee collimation system}

\author[a,1]{M. Behtouei,\note{Corresponding author.}}
\author[b,c]{E. Carideo}
\author[a]{M. Zobov}
\author[b,d]{M. Migliorati}
\affiliation[a]{INFN, Laboratori Nazionali di Frascati, P.O. Box 13, I-00054 Frascati, Italy}
\affiliation[b]{Dipartimento di Scienze di Base e Applicate per l'Ingegneria (SBAI), Sapienza University of Rome,\\Via Antonio Scarpa, 14, 00161 Rome, Italy}
\affiliation[c]{CERN, 1217 Meyrin, Geneva, Switzerland}
\affiliation[d]{INFN/Roma1, Istituto Nazionale di Fisica Nucleare, Piazzale Aldo Moro, 2, 00185, Rome, Italy }

\emailAdd{Mostafa.Behtouei@lnf.infn.it}

\abstract{The purpose of this paper is to calculate the longitudinal and transverse wakefields of the FCC collimators by using the electromagnetic codes ECHO3D and IW2D. We cross-checked our results using CST particle studio for long bunches, and found them to be in good agreement. The obtained results show that the collimators give one of the highest contributions to the overall FCC-ee wake potentials. Using the code PyHEADTAIL, we have found that the presence of the geometric wakefield of the collimators leads to the occurrence of transverse mode coupling instability (TMCI) at a significantly lower bunch population as compared to that of all other contributions and solutions to reduce this geometric term must be found. 
}


\begin{document}
\maketitle
\flushbottom

\section{Introduction}
 \label{sec:intro}
 
The FCC-ee is a proposed particle physics facility to be built in the CERN area \cite{ref1}, featuring a circular tunnel spanning 91 kilometers to house an advanced electron-positron collider.\\
As part of our efforts in the project, we are actively engaged in evaluating the impedance and collective effects associated with the collider \cite{refM1,refM2,refM3}. In particular we have focused our studies on the lowest energy machine (Z-pole)  since it is considered the most critical from the collective effects point of view. For this activity, we have recently found that one of the primary sources of impedance is represented is by the collimation system.\\
The FCC-ee machine will incorporate two types of collimators - beam halo collimators and SR collimators and masks. The former is used to ensure that the beam remains within well defined transverse area and is designed to intercept beam losses, while the latter will be used to intercept photons upstream of the interaction points (IPs). When designing beam halo collimators, robustness is a key criterion, particularly given the 20 MJ of stored beam energy in the machine. Tungsten collimators, which are susceptible to damage, are not an optimal option, as demonstrated by previous experience with SuperKEKB \cite{KEK}. Instead, low-Z materials provide greater robustness, but this comes at the expense of cleaning performance, necessitating the use of longer collimators and secondary collimation stages. On the other hand, the SR collimators and masks around the IPs must not have the same performances, they can  be made of tungsten and will be much shorter than the beam halo collimators.\\
To ensure proper beam halo collimation, movable, two-jaw collimators are foreseen with gaps set according to the transverse beam size and to the local betatron function $\beta_{(x/y)}$. These parameters, that is gap and $\beta_{(x/y)}$, have an important impact on the collimator impedance and, consequently, on the collective effects, and they must be properly optimized.\\
Indeed, the reduction of collimators' impedance in the FCC machine is an important issue for keeping under control the collective effects. Analytical estimations can offer valuable insights into how to reduce the geometrical impedance contribution. It is also crucial to include the resistive wall component in such an impedance reduction strategy, as it plays a significant role. Moreover, open questions remain regarding the optimal balance between reducing the local betatron function and the gap, as reducing the local betatron function can decrease the impedance weight but may require a reduction in the collimator gap, which approximately scales the impedance with the third power in the transverse plane. Further research and analysis are necessary to determine the best approach for reducing these impedance contributions and optimizing the machine's performance. \\
Studies on the impedance of the collimation system have been conducted in various accelerators, including LHC \cite{LHC1,LHC2} , DAFNE \cite{DAFNE}, KEKB \cite{KEKB}, SuperKEKB  \cite{SuperKEKB}, PEPII \cite{PEPII}, and LCLS \cite{LCLS}. In this paper we evaluate the wakefields produced by the collimators and their impact on the beam dynamics. In section 2 the analytical approach for evaluating the impedance and wakefield of a taper transition, as proposed in \cite{ref3,ref4}, is discussed in order to better understand the parameters that can be used to reduce the collimators' geometric contribution. Then, in section 3, we study both the geometric and resistive wall wakefields, also in comparison with the other sources, by means of the electromagnetic codes ECHO3D \cite{ref5} and IW2D \cite{refIW2D} and, in section 4,  their impact, in particular on the transverse beam dynamics, is studied with the help of the tracking code PyHEADTAIL \cite{refPy}.
\section{Analytical wakefield computation of taper transition}

Taper transition has been proposed as a potential solution for mitigating wakefield effects, particularly for transverse wakefields, as noted in \cite{ref2}. Yokoya in \cite{ref3} has developed equations for calculating the transverse and longitudinal impedances of a beam pipe with a slowly varying radius as a function of longitudinal position as 
\begin{equation}\label{2.1}
Z_{L}^{m=0}(k)=-\frac{i k Z_0}{4 \pi} I_{L}(a)
\end{equation}
\begin{equation}\label{2.2}
Z_{T}^{m} (k)=-\frac{i m Z_0}{(m+1) \pi} I^{(m)}_{T}(a)
\end{equation}
where  $I_{L}(a)=\int_{-\infty}^{\infty} dz (a')^2$ and $I^{(m)}_{T}(a)=\int_{-\infty}^{\infty} dz (\frac{a'}{a^m})^2 $. Equation \ref{2.1} gives the longitudinal impedance ($Z_L$) for a beam pipe with a slowly varying radius, where $k$ is the wave number, $Z_0$ is the impedance of free space and $a'$ is the gradient of the taper radius with respect to the longitudinal coordinate z. Equation \ref{2.2} gives the corresponding transverse impedance ($Z_T$) for the azimuthal mode number $m$. Stupakov extended Yokota's methods to calculate the low-frequency impedance for any shape of the transition cross-section. For the case of an axisymmetric transition, the following equations are provided \cite{ref4}
\begin{equation}
Z_{L}^{m=0} (k)=\frac{Z_0}{2\pi}  ln \frac{a_1}{a_2}-i\frac{kZ_0}{4\pi} I_{L}(a)
\end{equation}
\begin{equation}
Z_{T}^{dipole} (k)=\frac{ Z_0}{2k \pi} {\biggl (}\frac{1}{a_1^2} - \frac{1}{a_2^2}{\biggl)}- i \frac{  Z_0}{2\pi} I^{(1)}_{T}(a).
\end{equation}
In the case of identical pipes ($a_1=a_2$), the first term of both the longitudinal and dipolar impedances vanishes, as the logarithmic term in the longitudinal impedance and the difference one in the inverse square terms in the dipolar impedance cancel out. Therefore, in this case, the impedances become purely inductive and are solely dependent on the integrals $I_L(a)$ and $I^{(1)}_T(a)$ respectively. \\
To calculate the wakefield, we can use the definition of inverse Fourier transform as follows:
\begin{equation}
f(z)=\frac{1}{2\pi} \int_{-\infty}^{\infty} \tilde{f}(k) e^{ikz} dk,
\end{equation}
where $\tilde{f}(k)$ is the Fourier transform of $f(z)$.\\
Applying this definition to $Z_{L}^{m=0} (k)$, we get
\begin{equation}\label{2.5}
W_{||} (z)=\mathscr{F}^{-1}{\biggl [}-i\frac{kZ_0}{4\pi} \int   (a')^2  dz{\biggl ]} = -\frac{ Z_0}{2\sqrt{2\pi}} \delta'(z) \int   (a')^2  dz.
\end{equation}
where $\delta'(z)$ is the derivative of the Dirac delta function. To solve the integral above, we used the identity
\begin{equation}
\int_{-\infty}^{\infty} e^{ikz} dk=2\pi \delta(z),
\end{equation}
where $\delta(z)$ is the Dirac delta function. \\
By convolving this wakefield with a Gaussian distribution $\lambda (z)=\frac{q}{\sqrt{2\pi} \sigma_z} e^{- \frac{z^2}{2\sigma_z^2}} $, we can obtain the wake potential for a given bunch length $\sigma_z$ as \cite{Zobov}
\begin{figure}[b]
 \begin{center}
  \fbox{   \includegraphics[width= 0.3\textwidth ]{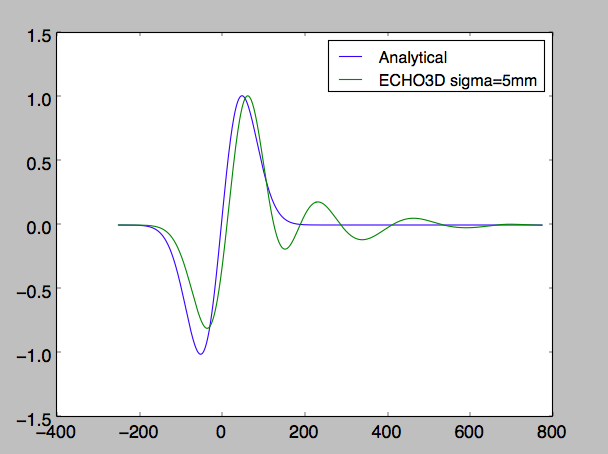}}
\caption{Comparison of analytical expression and numerical simulations for the wake potential of a 5 mm Gaussian bunch passing through a collimator with a half gap of 3.5 mm, using ECHO3D.}
\end{center}
\end{figure}
\begin{equation}\label{2.6}
W_{||}=\frac{1}{q} \int_{-\infty}^\infty \omega_{||}(z-z^,) \lambda(z^,) dz^,
\end{equation}
where q is the total bunch charge. Substituting the wakefield given by Eq. (\ref{2.5}) into Eq. (\ref{2.6}), we obtain an expression for $W_{||}$ in terms of the collimator jaws slope $a'$ and the length of the taper $L$ as

\begin{equation}
W_{||}=-\frac{1}{q} \int_{-\infty}^\infty [{-\frac{Z_0}{2\sqrt{2\pi}} \delta'(z-z^,)  (a')^2 }] \lambda(z^,) dz^,=-Z_0 L a'^2\frac{  z}{4\pi \sigma_z^3} \ e^{- \frac{z^2}{2\sigma_z^2}} 
\end{equation}
The same method can be used also for the transverse plane wake potential, which gives

\begin{equation}
W_{T}=-\frac{1}{q} \int_{-\infty}^\infty [{-\frac{Z_0}{\sqrt{2\pi}} \delta (z-z^,) \int   \frac{(a')^2}{a^2}  dz}] \lambda(z^,) dz^,= \frac{  Z_0 }{2\pi \sigma_z} \  e^{- \frac{z^2}{2\sigma_z^2}}  \int   \frac{a'^2}{a^2}  dz.
\end{equation}
To illustrate the validity of the above equations, Fig. (1) shows a comparison between the analytical expression of the longitudinal wake potential and a simulation using ECHO3D for a 0.4 mm Gaussian bunch passing through a collimator with a half gap of 3.5 mm. The initial beam pipe radius is of 35 mm. The length of the taper is 75 mm. The plot shows the results of the comparison, highlighting the close agreement between the two methods. Expressions (2.9) and (2.10) can be used to understand the dependence of the geometrical contribution to the wakefield of a collimator with respect to the length of the taper in under to reduce this wakefield source.

\section{Numerical solution using CST and ECHO3D}
Figure 2(a) shows the SuperKEKB collimator original shape  which was the starting point of our design for the FCC collimator model \cite {refT}. The impedance model for FCC-ee is regularly updated to reflect the evolving parameters and information provided by different groups. In Table (1) we show the main parameters of the betatron and off-momentum collimators \cite{refAndrey} foreseen for the lowest energy machine with 4 interaction points (IPs). Figure 2(c) shows that the adapted shape of the FCC collimator is considerably longer than that of the SuperKEKB model to compensate for the loss of cleaning performance \cite{ref7,ref8}.\\
It has been decided to employ ECHO3D \cite{ref5} for most of our numerical simulations since the code is particularly suitable for calculation of wakefields of short bunches traveling in long tapered structures. Due to a low-dispersive numerical technique used in the simulations, ECHO3D provides a fast convergence of the results even with a coarse mesh, thus helping to save substantially the required computational time.
  \begin{figure}[b]
 \begin{center}
  \fbox{  \includegraphics[width=0.32 \textwidth ]{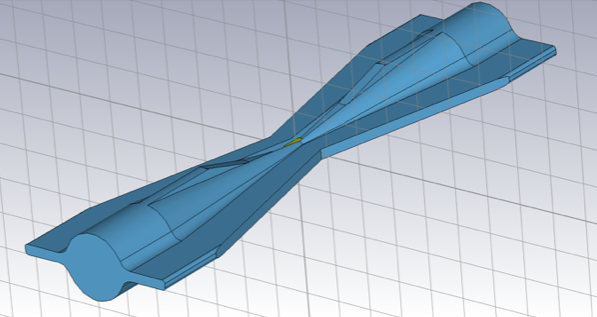}}(a)
    \fbox{   \includegraphics[width= 0.55\textwidth ]{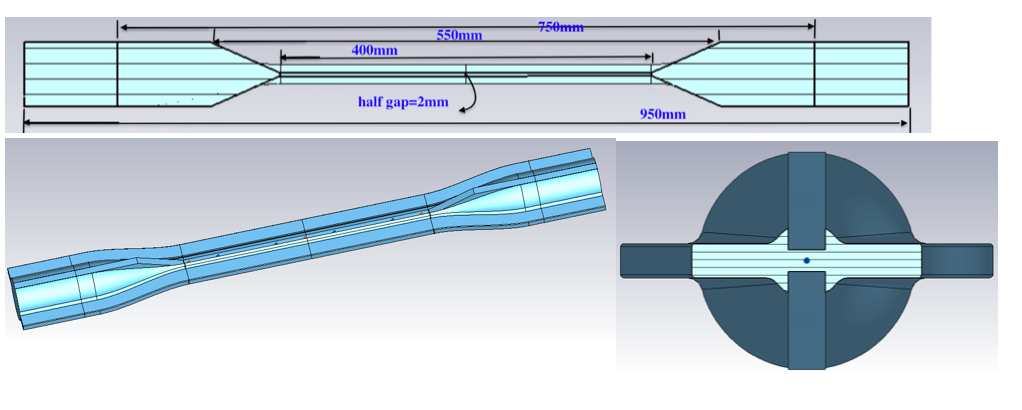}}(c)
\end{center}
\caption{a) The original shape of the collimator used in SuperKEKB \cite{refT}. b) A proposed FCC-ee collimator model. It should be noted that the length of these collimators is much longer than the SuperKEKB model, ranging from 0.3 to 0.4 meters instead of just a few millimeters. }
\end{figure}
\begin{table}[h]
\caption{ Summary of the collimator settings for the Z-pole and for the 4 IPs layout has shown in Table (1). The table includes both betatron and off-momentum collimators. The synchrotron collimators and masks upstream the IP are not included in the table. The length of these collimators is much longer than the SuperKEKB model (0.3 - 0.4 m instead of few mm).}
 \begin{center}
{\tiny{\begin{tabular}{||c| c| c| c| c| c|  c| c| c| c| c| c| c| c||} 
\hline
name&type &$\ell$ [m]& nsigma&half-gap [m]&material&plane&angle[deg]&offset$_x$ [m]&offset$_y$[m]&beta$_x$[m]&beta$_y$[m]\\ 
\hline\hline
tcp.h.b1& primary &0.4&11.0&0.005504&MoGR&H&0.0&0.0&0.0&352.578471&113.054110\\ 
\hline
tcp.v.b1& primary &0.4& 65.0&0.002332&MoGR&V&90.0&0.0&0.0&147.026106&906.282898\\  
\hline
tcs.h1.b1 & secondary& 0.3&13.0&0.004162&Mo&H&0.0&0.0&0.0&144.372060&936.118623\\ 
\hline
tcs.v1.b1&secondary &0.3&75.5&0.00203&Mo&V&90.0&0.0&0.0&353.434125&509.320452\\  
 \hline
 tcs.h2.b1& secondary  &0.3&13.0&0.005956&Mo&H&0.0&0.0&0.0&295.623450&1419.375106\\ 
 \hline
  tcs.v2.b1&  secondary &  0.3&75.5&0.002116&Mo&V&90.0&0.0&0.0&494.235759&554.055888\\ 
  \hline
tcp.hp.b1& primary & 0.4& 29.0&0.005755&MoGR&H&0.0&0.0&0.0&55.469637&995.306256\\ 
\hline
tcs.hp1.b1&secondary&0.3&  32.0&0.01649&Mo&H&0.0&0.0&0.0&373.994993&377.277726\\ 
\hline
tcs.hp2.b1&secondary&0.3&  32.0&0.011597&Mo&H&0.0&0.0&0.0&184.970621&953.229862\\ 
\hline
\end{tabular}}}
\end{center}
\end{table}
\begin{figure}
 \begin{center}
   \fbox{  \includegraphics[width=0.35 \textwidth ]{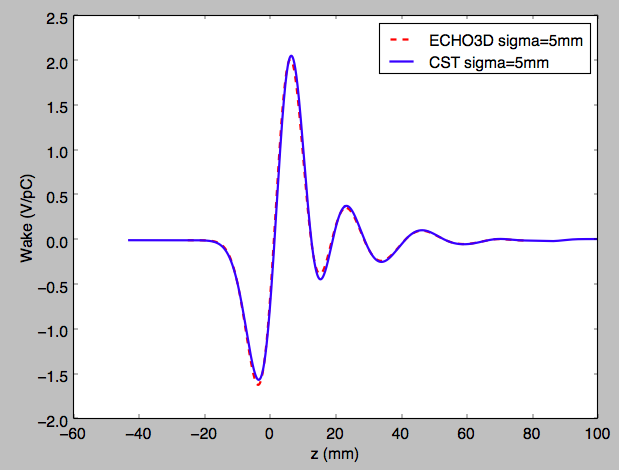}}(a)
 \fbox{  \includegraphics[width=0.35\textwidth ]{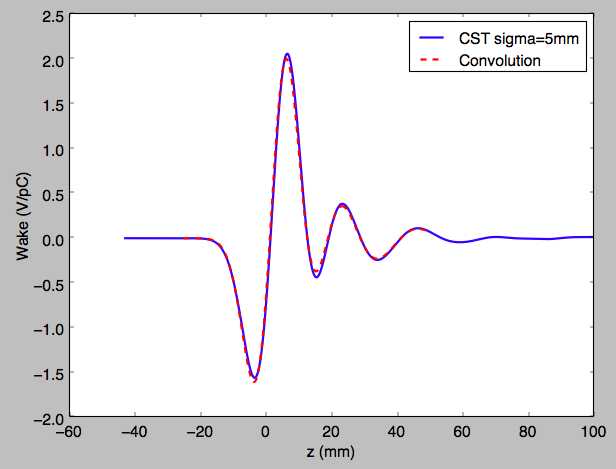}}(b)
  \fbox{  \includegraphics[width=0.35 \textwidth ]{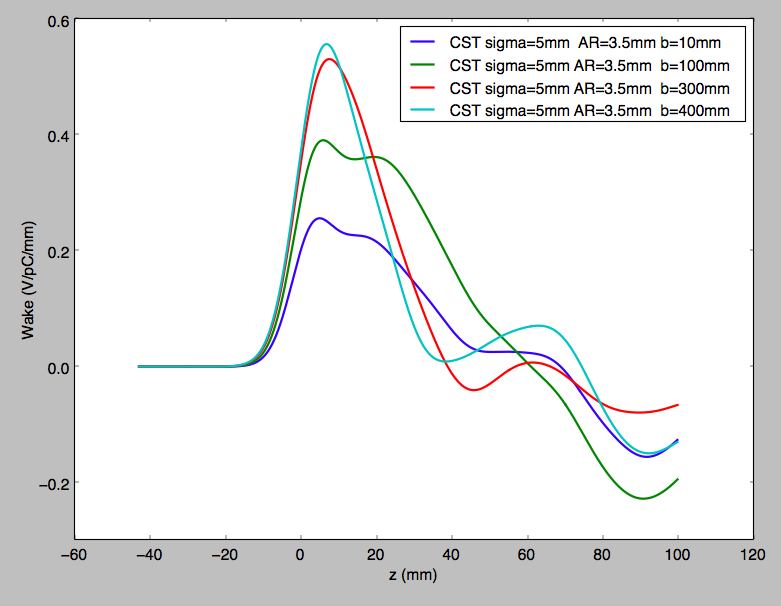}}(c)
   \fbox{   \includegraphics[width= 0.35\textwidth ]{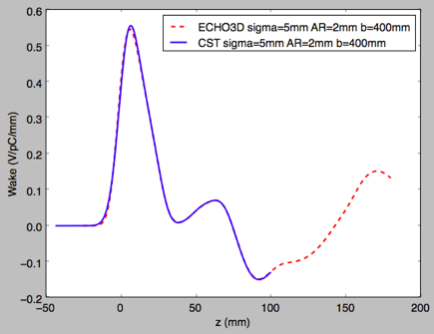}}(d)
\end{center}
\caption{a) Comparison of the geometric longitudinal wake potential of a 5 mm Gaussian bunch computed using ECHO3D and CST. b) The convolution method used to obtain the wake potential of a desired sigma. c) The wake potential is almost identical between 300 mm and 400 mm for collimators with a length of 0.3 - 0.4 m. d) The transverse dipolar wake potential of a 0.4 mm Gaussian bunch computed using ECHO3D ( high number of mesh cells required by CST). }
\end{figure}
For investigation of collective effects in FCC-ee we evaluate the wake potentials of a 0.4 mm Gaussian bunch used as a psedo-Green function for beam dynamics studies. This bunch length is about 10 times shorter than the natural one. The wake potential calculation for such short bunches (0.4 mm) by using other numerical codes than ECHO3D would require excessively high number of mesh cells.
\begin{figure}[t]
 \begin{center}
 {\color{red}  }
           \fbox{  \includegraphics[width=0.26\textwidth ]{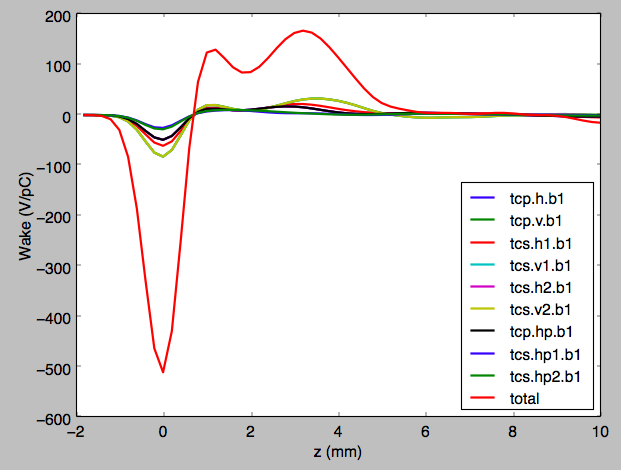}} (a)
         \fbox{  \includegraphics[width=0.26\textwidth ]{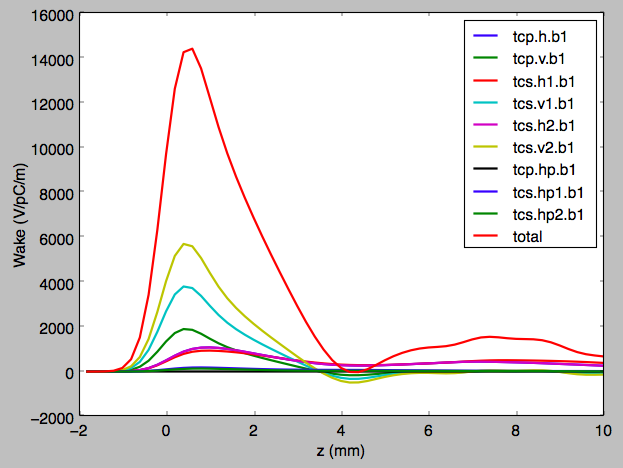}} (b) 
         \fbox{  \includegraphics[width=0.24\textwidth ]{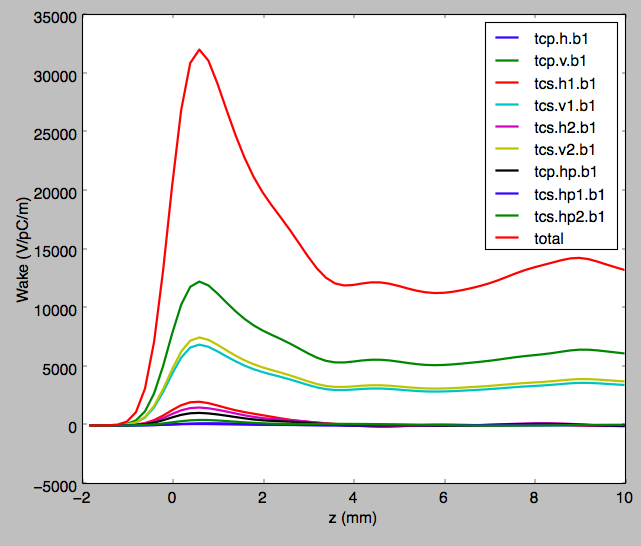}} (c)
         \fbox{  \includegraphics[width=0.26\textwidth ]{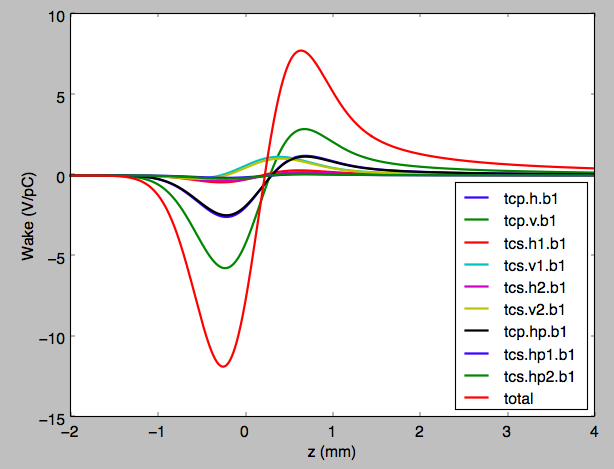}} (d) 
         \fbox{  \includegraphics[width=0.26\textwidth ]{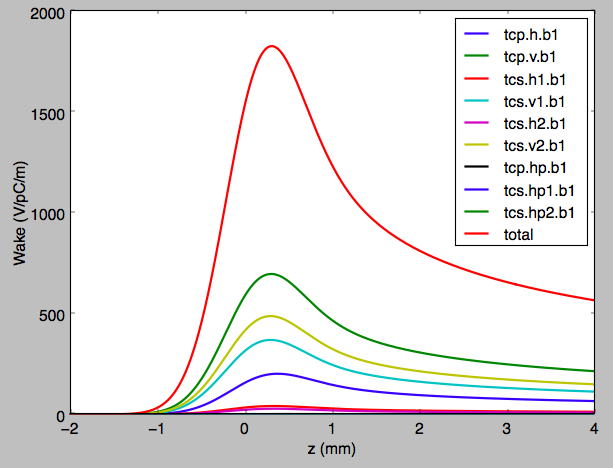}} (e)
         \fbox{  \includegraphics[width=0.26\textwidth ]{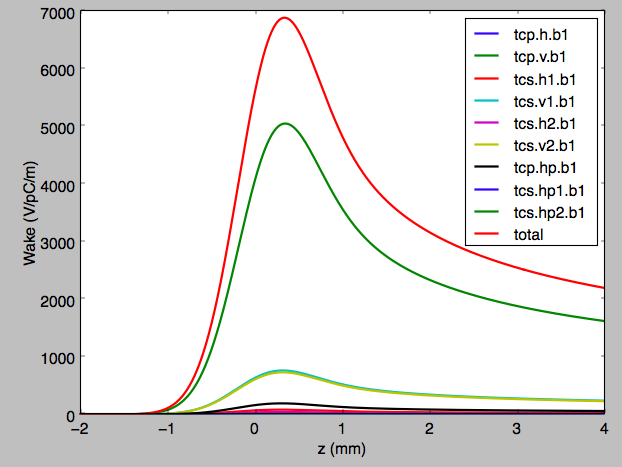}} (f) 
          \fbox{  \includegraphics[width=0.26\textwidth ]{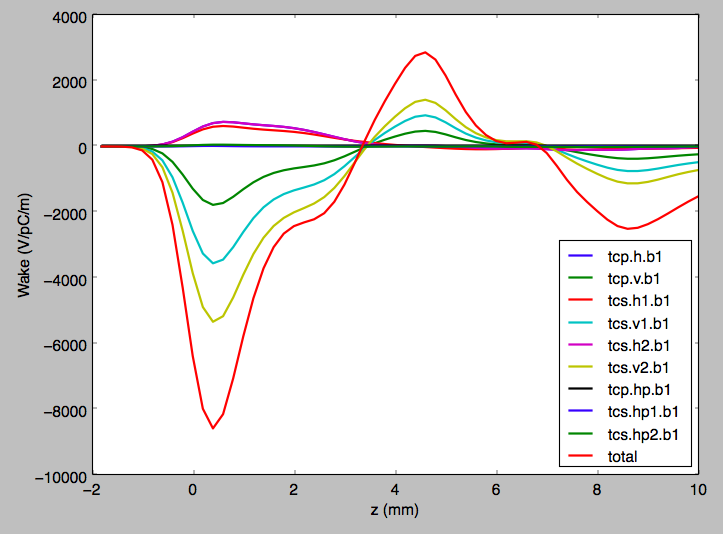}} (g)
         \fbox{  \includegraphics[width=0.26\textwidth ]{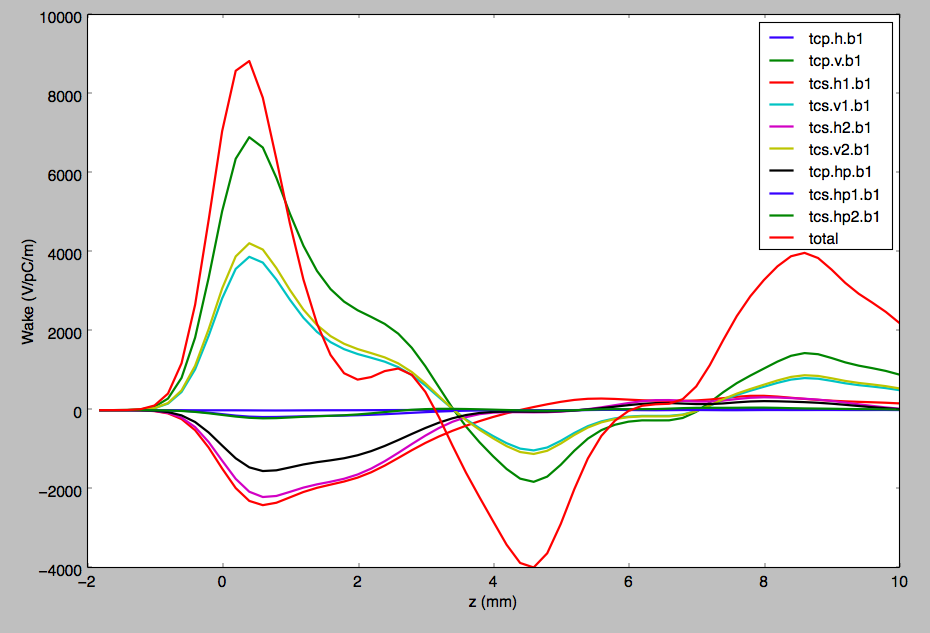}} (h) 
         \fbox{  \includegraphics[width=0.26\textwidth ]{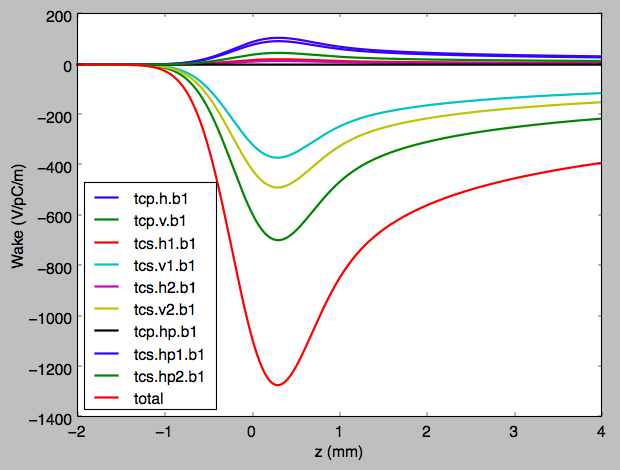}} (i)
         \fbox{  \includegraphics[width=0.26\textwidth ]{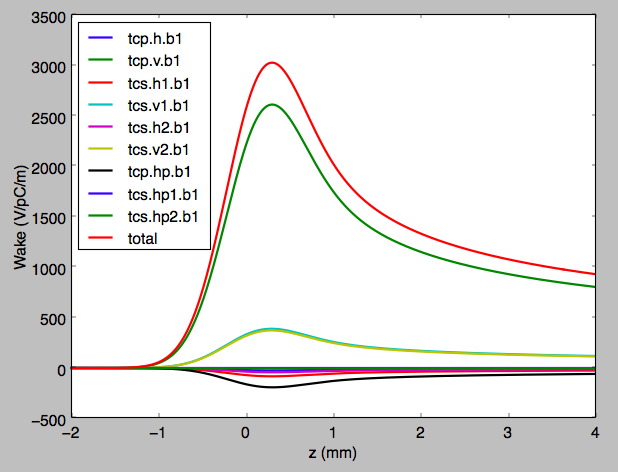}} (j) 
         \caption{ The longitudinal (a), transverse horizontal dipolar (b), transverse vertical dipolar (c), transverse horizontal quadrupolar (g) and transverse vertical quadrupolar (h) geometrical wake potential of collimators for a 0.4 mm Gaussian bunch calculated using ECHO3D. Similarly, the longitudinal (d), transverse horizontal dipolar (e), transverse vertical dipolar (f), transverse horizontal quadrupolar (i) and transverse vertical quadrupolar (j) components of the resistive wall  wake potential for the same 0.4 mm bunch computed using IW2D. }
   \end{center}
\end{figure} 
However, in order to validate the ECHO3D results, CST Particle Studio  \cite{ref6} can be used with longer bunches. As an example, Fig. 3 (a) shows a comparison of the longitudinal wake potential of a 5 mm Gaussian bunch computed using ECHO3D and CST. Additionally, CST result can be compared with the convoluted wake potential obtained for the quasi-Green function calculated by ECHO3D (see Fig 3 (b)). In both cases the agreement is excellent. Besides, we see that wake of the 0.4 mm Gaussian bunch can be safely applied as a Green function to reproduce the wake potentials for typical bunch lengths of in FCC-ee.\\
We have also calculated the transverse dipolar wake potentials of a 0.4 mm Gaussian bunch for the collimators with different length of the jaws. From Fig. 3(c), it can be observed that the wake potential is almost the same for collimator lengths ranging from 300 mm to 400 mm. Figure 3(d) shows the validation of the result for a 400 mm collimator length. To simulate the transverse dipolar wake potential of a 0.4 mm Gaussian bunch with different collimator gaps, we fixed the length of the jaws to 400 mm.\\ 
In addition to the geometric contribution, also the resistive wall term, produced by the finite conductivity of the pipe walls plays an important role. To evaluate the resistive wall contribution to the wakefield of the collimators, we used IW2D \cite{refIW2D} and simulated parallel plates with infinite thickness.\\
Separating the resistive wall (RW) and geometric contributions is essential to understand the origin and relative importance of each contribution to the overall impedance of the collimators because we can identify which one is the dominant source of impedance and develop mitigation methods accordingly.
\begin{figure}[t]
 \begin{center}
 {\color{red}  }
         \fbox{  \includegraphics[width=0.34\textwidth ]{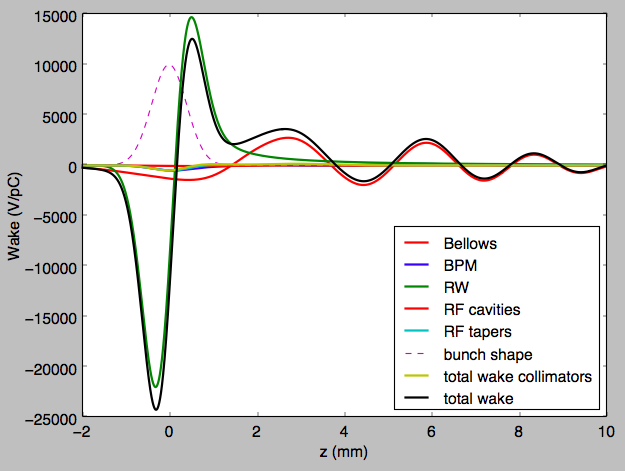}} (a)
         \fbox{  \includegraphics[width=0.34\textwidth ]{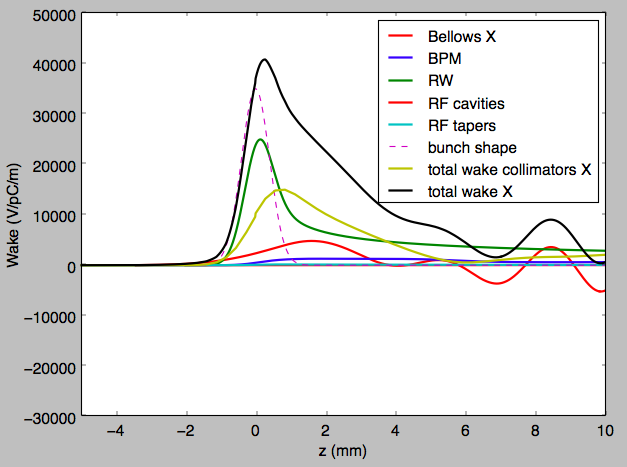}} (b) 
         \fbox{  \includegraphics[width=0.34\textwidth ]{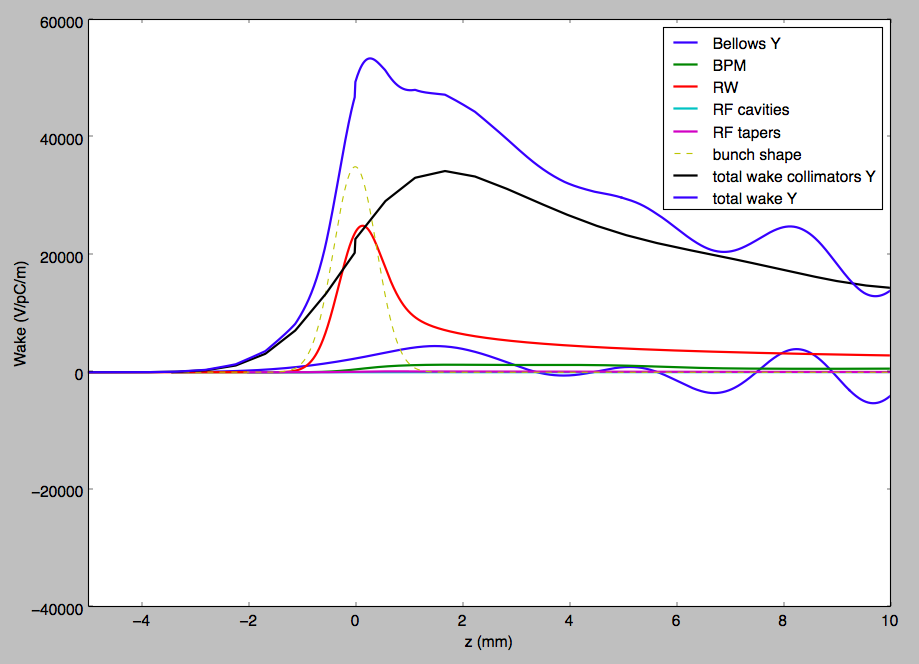}} (c) 
         \caption{ a) The total longitudinal b) horizontal dipolar and c) vertical dipolar wake potential of the FCC machine for a 0.4 mm Gaussian bunch.}
   \end{center}
\end{figure} 
The Figs. 4 represent various types of wake potentials of collimators computed using different computational tools for a 0.4 mm Gaussian bunch, ECHO3D for the geometrical term of the collimators and IW2D for the resistive wall wake potentials. These figures provide a comprehensive overview of the wake potentials of collimators, including their geometric and resistive wall dipolar and quadrupolar components. It's worth emphasizing that the transverse plane, specifically the vertical dipolar component of the geometric wake potential as shown in Fig (3c), carry the most significant wake potential and therefore should be considered the most critical. Based on the identification of the vertical collimators with the half gap of approximately 2 mm (tcp.v.b1, tcs.v1.b1, and tcs.v2.b1 as illustrated in Figs. (4c) and (4d)), we have determined that these collimators have the greatest contribution to the geometric wakefield.\\
The wake potentials that have been evaluated so far consider several components, including the resistive wall, bellows, beam position monitors (BPM), and the RF system, which includes tapers connecting the cryo-modules in both longitudinal and transverse planes. These contributions are illustrated in Figs. 5. It is worth noting that, apart from the resistive wall and the bellows, the other devices have a negligible impact on the longitudinal wake potentials. However, by comparing the plots of Figs. 4 and 5, we can observe that, in particular in the transverse plane, the geometric contribution of collimators is almost equal to the total of all the other wakefield sources evaluated so far.

\section{Collective effects}
The simulation code PyHEADTAIL can analyze the combined effect of longitudinal and transverse wakefields in a self-consistent manner (excluding the beam-beam effects). This interaction between the two planes resulted particularly important for FCC-ee \cite{refM2}, for which it has been demonstrated that the transverse mode coupling instability (TMCI) threshod is affected by the longitudinal wakefield. Moreover, also a bunch by bunch feedback system, necessary to damp the transverse coupled bunch instability              due to the real part of the transverse impedance, represents an important tool to mitigate the TMCI. The results of this analysis using a strong feedback system with a damping time of 4 turns and with the pseudo-Green function including all the contributions of wake potentials except the geometrical one due to the collimators is shown in Fig. 6, left-hand side, where we have represented the real part of the tune shift for the first azimuthal transverse coherent oscillation modes as a function of the bunch population, normalized by the synchrotron tune $Q_{s0}$ \cite{refM4}. The results in the right hand side of the figure shows the data with the influence of the collimator's geometric wakefield. We observed in both cases that there is '-1' mode instability, but no shift of the 0 mode thanks to the feedback system.\\
\begin{figure}[h]
 \begin{center}

 {\color{red}  }
         \fbox{  \includegraphics[width=0.49\textwidth ]{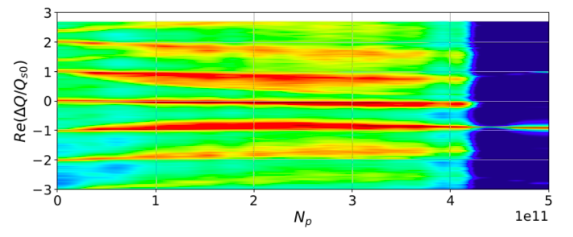}} 
         \fbox{  \includegraphics[width=0.43\textwidth ]{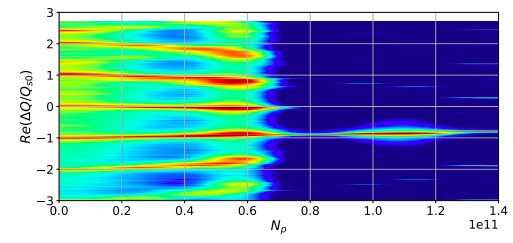}} 
         \caption{ Real part of the tune shift of the first azimuthal transverse coherent oscillation modes normalised by the synchrotron tune $Q_{s0}$ as a function of bunch population without (left) and with (right) the effect of geometric wakefield  of the collimators.  .}
   \end{center}
\end{figure} 
From the figure we can see that in the presence of the geometric wakefield of the collimators, the transverse mode coupling instability (TMCI) arises when the bunch population is less than $N_p=1\times 10^{11}$. In contrast, when only the RW contribution of the collimators is considered, TMCI occurs at a higher bunch population of approximately $N_p=4.2 \times 10^{11}$.\\
As expected, the geometric wakefield due to the collimators reduces the threshold by a very large amount and solutions to decrease the amplitude of such wakes must be found. According to the studies of section 2, we are currently exploring the maximum length of tapers that can be implemented to effectively reduce the impact of the geometric wakefield. \\
In addition, non-linear tapers \cite{refNLT} and non-linear collimation optics \cite{refNLO}  can also be explored as potential methods for mitigating the effects of the geometrical impedance. With the non-linear optics solution, also the RW conribution of the collimators can be effectively reduced. These approaches may offer additional advantages over linear tapers, such as greater flexibility and control over the beam dynamics. Our investigation into this matter is ongoing.
 \section{Summary}
Similarly to other particle colliders, wakefields of collimators can substantially affect the beam dynamics and respectively performance of FCC-ee. In this paper we study the wakefields of the betatron and off-momentum collimators set for the Z mode of collider operation with 4 interaction points.\\
For numerical simulations of the collimator geometric wakes the two different codes, CST Particle Studio and ECHO3D, are used. In this way we can cross-benchmark the obtained results. In order to account for the resistive wall contribution, the parallel plate approximation is assumed and IW2D code is used for the wakefield evaluation.\\
In order to validate the results we first compare the wakes obtained numerically with the available analytical expressions revealing a good agreement (Fig.1). Then, the convolution method is used to compare the collimator geometric wakes potentials computed by using CST and ECHO3D. The results are in a perfect agreement as shown in Fig. 3.\\
Finally, by summing up the contributions of all collimators, it has been found that the total collimators wake is very high. In particular, the vertical transverse dipolar wake potential weighted by the respective beta functions at the collimator locations is comparable to the overall wake contribution coming from all other vacuum chamber components considered so far.\\
In order to evaluate the impact of the collimator wake on the beam dynamics in FCC-ee, PyHEADTAIL code has been used to simulate the transverse mode coupling instability (TMCI) in FCC-ee. In particular, it has been demonstrated that even with the presence of a strong feedback, an addition of the collimator geometric wakefields leads to TMCI occurring at a significantly lower threshold with respect to the nominal bunch population.\\ 
In conclusion, our analysis indicates that the contribution of collimators into the FCC impedance/wake budget is significant so that dedicated measures should be taken to mitigate their harmful impact on the collider performance. 
\section{Acknowledgement}
This project is supported from the European Union's Horizon 2020 research and innovation programme under grant agreement No 951754 and partially by INFN National committee V through the ARYA project. The authors express their gratitude to M. Boscolo, F. Zimmermann and A. Gallo for their constant support, as well as the CERN collimation system group, with special thanks to A. Andrey for providing the collimation system table.

\end{document}